\documentclass[conference]{IEEEtran} 
\IEEEoverridecommandlockouts
\usepackage{cite}
\usepackage{amsmath,amssymb,amsfonts}
\usepackage{algorithmic}
\usepackage{graphicx}
\usepackage{textcomp}
\usepackage{xcolor}
\def\BibTeX{{\rm B\kern-.05em{\sc i\kern-.025em b}\kern-.08em
    T\kern-.1667em\lower.7ex\hbox{E}\kern-.125emX}}
\begin{document}

\title{Deployment of Real-Time Network Traffic Analysis \\ using GraphBLAS Hypersparse Matrices \\ and D4M Associative Arrays
\thanks{
This material is based upon work supported by the Assistant Secretary of Defense for Research and Engineering under Air Force Contract No. FA8702-15-D-0001. Any opinions, findings, conclusions or recommendations expressed in this material are those of the author(s) and do not necessarily reflect the views of the Assistant Secretary of Defense for Research and Engineering. Research was also sponsored by the United States Air Force Research Laboratory and the Department of the Air Force Artificial Intelligence Accelerator and was accomplished under Cooperative Agreement Number FA8750-19-2-1000. The views and conclusions contained in this document are those of the authors and should not be interpreted as representing the official policies, either expressed or implied, of the Department of the Air Force or the U.S. Government. The U.S. Government is authorized to reproduce and distribute reprints for Government purposes notwithstanding any copyright notation herein.
}
}

\author{\IEEEauthorblockN{Michael Jones$^1$, Jeremy Kepner$^1$, Andrew Prout$^1$, Timothy Davis$^2$, William Arcand$^1$, David Bestor$^1$, \\ William Bergeron$^1$,   Chansup Byun$^1$, Vijay Gadepally$^1$,  Micheal Houle$^1$, Matthew Hubbell$^1$,  Hayden Jananthan$^1$,  \\  Anna Klein$^1$, Lauren Milechin$^1$, Guillermo Morales$^1$, Julie Mullen$^1$, Ritesh Patel$^1$, Sandeep Pisharody$^1$, \\  Albert Reuther$^1$, Antonio Rosa$^1$, Siddharth Samsi$^1$,  Charles Yee$^1$, Peter Michaleas$^1$
\\
\IEEEauthorblockA{$^1$MIT, $^2$Texas A\&M}}}
\maketitle

\begin{abstract}
Matrix/array analysis of networks can provide significant insight into their behavior and aid in their operation and protection. Prior work has demonstrated the analytic, performance, and compression capabilities of GraphBLAS (graphblas.org) hypersparse matrices and D4M (d4m.mit.edu) associative arrays (a mathematical superset of matrices). Obtaining the benefits of these capabilities requires integrating them into operational systems, which comes with its own unique challenges. This paper describes two examples of real-time operational implementations. First, is an operational GraphBLAS implementation that constructs anonymized hypersparse matrices on a high-bandwidth network tap. Second, is an operational D4M implementation that analyzes daily cloud gateway logs. The architectures of these implementations are presented. Detailed measurements of the resources and the performance are collected and analyzed. The implementations are capable of meeting their operational requirements using modest computational resources (a couple of processing cores). GraphBLAS is well-suited for low-level analysis of high-bandwidth connections with relatively structured network data. D4M is well-suited for higher-level analysis of more unstructured data. This work demonstrates that these technologies can be implemented in operational settings.
\end{abstract}


\section{Introduction}
 
Matrix/array-based analysis of graphs and networks can provide significant insight into their behavior and aid in their operation and protection \cite{soule2004identify, zhang2005estimating, mucha2010community, tune2013internet, do20classifying, weed2022beyond, kawaminami2022enrichment, rodriguez2022arachne, durbeck2022kalman, sathre2022jaccard, zhao2022hugraph, zhang2022sharp, mandulak2022explicit, kang2022analyzing, moe2022implementing, singh2022efficient, liu2022nwhy, acosta2022families, osama2022essentials}.  Extensive prior work has demonstrated the analytic, performance, and compression capabilities of the open standard GraphBLAS -- Graph Basic Linear Algebra Subprograms -- (graphblas.org) hypersparse matrices and D4M  -- Dynamic Distributed Dimensional Data Model --  (d4m.mit.edu) associative arrays (a mathematical superset of matrices)\cite{buluc17design, yang2018implementing, kepner2018mathematics, davis2019algorithm, mattson2019lagraph, davis2019write, aznaveh2020parallel, brock2021introduction, davis2021algorithm, pelletier2021graphblas, jones2022graphblas, trigg2022hypersparse, jananthan2022pyd4m, afanasyev2022graphblas, mastoras2022nonblocking, costanza2022towards, brock2022graphblas, scolari2023effective}.

\begin{figure}
\center{\includegraphics[width=1.0\columnwidth]{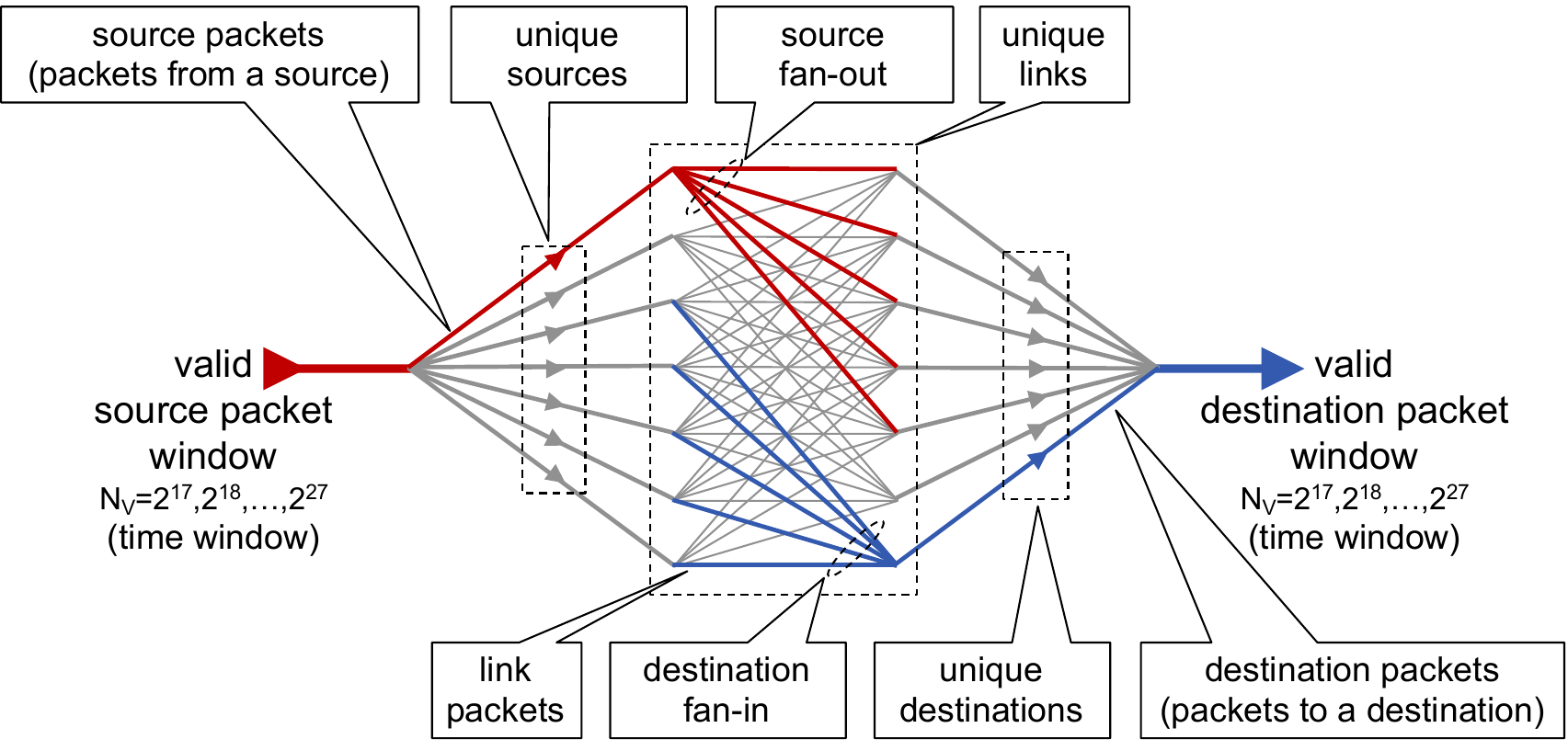}}
      	\caption{{\bf Generic Streaming Network Traffic Quantities.} Network traffic streams of $N_V$ valid packets are divided into a variety of quantities for analysis: source packets, source fan-out, unique source-destination pair packets (or links), destination fan-in, and destination packets.  Figure adapted from \cite{kepner19hypersparse}.}
      	\label{fig:NetworkDistribution}
\end{figure}

\begin{table}
\caption{Network Quantities from Traffic Matrices/Arrays}
\vspace{-0.25cm}
Formulas for computing network quantities from  traffic matrix/array ${\bf A}_t$ at time $t$ in both summation and matrix notation. ${\bf 1}$ is a column vector of all 1's, $^{\sf T}$  is the transpose operation, and $|~|_0$ is the zero-norm that sets each nonzero value of its argument to 1\cite{karvanen2003measuring}.  These formulas are unaffected by permutations and will work on anonymized data.  Table adapted from \cite{kepner2020multi}.
\begin{center}
\begin{tabular}{p{1.1in}p{1.0in}p{0.7in}}
\hline
{\bf Aggregate} & {\bf ~~~~Summation} & {\bf Matrix/Array} \\
{\bf Property} & {\bf ~~~~~~Notation} & {\bf ~~~Notation} \\
\hline
Valid packets $N_V$ & $~\sum_i ~ \sum_j ~ {\bf A}_t(i,j)$ & $~{\bf 1}^{\sf T} {\bf A}_t {\bf 1}$ \\
Unique links & $~~\sum_i ~ \sum_j |{\bf A}_t(i,j)|_0$  & ${\bf 1}^{\sf T}|{\bf A}_t|_0 {\bf 1}$ \\
Link packets from $i$ to $j$ & $~~~~~~~~~~~~~~{\bf A}_t(i,j)$ & ~~~$~{\bf A}_t$ \\
Max link packets & $~~~~~\max_{ij}{\bf A}_t(i,j)$ & $\max({\bf A}_t)$ \\
\hline
Unique sources & $~\sum_i |\sum_j ~ {\bf A}_t(i,j)|_0$  & ${\bf 1}^{\sf T}|{\bf A}_t {\bf 1}|_0$ \\
Packets from source $i$ & $~~~~~~~\sum_j ~ {\bf A}_t(i,j)$ & ~~$~~{\bf A}_t  {\bf 1}$ \\
Max source packets  & $ \max_i \sum_j ~ {\bf A}_t(i,j)$ & $\max({\bf A}_t {\bf 1})$ \\
Source fan-out from $i$ & $~~~~~~~~~~\sum_j |{\bf A}_t(i,j)|_0$  & ~~~$|{\bf A}_t|_0 {\bf 1}$ \\
Max source fan-out & $ \max_i \sum_j |{\bf A}_t(i,j)|_0$  & $\max(|{\bf A}_t|_0 {\bf 1})$ \\
\hline
Unique destinations & $~\sum_j |\sum_i ~ {\bf A}_t(i,j)|_0$ & $|{\bf 1}^{\sf T} {\bf A}_t|_0 {\bf 1}$ \\
Destination packets to $j$ & $~~~~~~~\sum_i ~ {\bf A}_t(i,j)$ & ${\bf 1}^{\sf T}|{\bf A}_t|_0$ \\
Max destination packets & $ \max_j \sum_i ~ {\bf A}_t(i,j)$ & $\max({\bf 1}^{\sf T}|{\bf A}_t|_0)$ \\
Destination fan-in to $j$ & $~~~~~~~~~~\sum_i |{\bf A}_t(i,j)|_0$ & ${\bf 1}^{\sf T}~{\bf A}_t$ \\
Max destination fan-in  & $ \max_j \sum_i |{\bf A}_t(i,j)|_0$ & $\max({\bf 1}^{\sf T}~{\bf A}_t)$ \\
\hline
\end{tabular}
\end{center}
\label{tab:Aggregates}
\end{table}%

A primary benefit of matrix/array-based analysis of graphs and networks is the efficient computation of a wide range of analytics quantities via simple mathematical formulas.  Figure~\ref{fig:NetworkDistribution} illustrates essential quantities found in all streaming dynamic networks. In real-world applications of streaming graph/network data, it is common to filter the events/packets to a valid set for any particular analysis.   Such filters may limit particular sources, destinations, protocols, and time windows.  At a given time $t$, $N_V$ consecutive valid events/packets are aggregated from the traffic into a hypersparse matrix or associative array ${\bf A}_t$, where ${\bf A}_t(i,j)$ is the number of valid events/packets between the source $i$ and destination $j$. The sum of all the entries in ${\bf A}_t$ is equal to $N_V$
$$
    \sum_{i,j} {\bf A}_t(i,j) = N_V
$$
All the network quantities depicted in Figure~\ref{fig:NetworkDistribution} can be readily computed from ${\bf A}_t$ using the formulas listed in Table~\ref{tab:Aggregates}.  If the events/packets are mostly unique, an $N{\times}N$ matrix with a given number of events/packets $N_V$ can be denoted as dense ($N_V \sim N^2$), sparse ($N_V \sim N$), or hypersparse ($N_V \ll N$) \cite{bulucc2009parallel}.  Associative arrays are a superset of matrices and extend the index and value sets of matrices to be any strict totally orderable set, which allows using unstructured data, such as strings, as either indices or values.  By design, associative arrays are almost always hypersparse.  Mathematically, both hypersparse matrices and associative arrays are defined as mappings from row keys $I$ and column keys $J$ to a value set $V$
$$
    {\bf A}: I{\times}J \rightarrow V
$$
In this context,  the hypersparse matrices have $I = J = \{0,\dots,2^{32}-1\}$;  $V$ are real numbers approximated as double precision floating point values.  Likewise, the associative arrays have $I = J = V$ that are all strict totally ordered sets approximated as strings.

The ability to handle hypersparse and/or unstructured data allows simple algebraic equations to be used to analyze streaming network data.  Obtaining the benefits of these capabilities in real-world applications requires integrating them into operational systems, which comes with its own unique challenges.  This paper describes two such instances of real-time operational implementations as follows.  First, building on prior work \cite{jones2022graphblas}, an operational GraphBLAS implementation that constructs anonymized hypersparse matrices on a high-bandwidth network tap.  Second, an operational D4M implementation that analyzes daily cloud gateway logs.  The architectures of these implementations are presented.  Detailed measurements of the resources and the performance are collected and analyzed.    The goal is to illustrate to others some effective architectures for deploying matrix/array-based analysis on streaming networks and to set the expectations on the potential rates and resources required.

\begin{figure}
\center{\includegraphics[width=1.0\columnwidth]{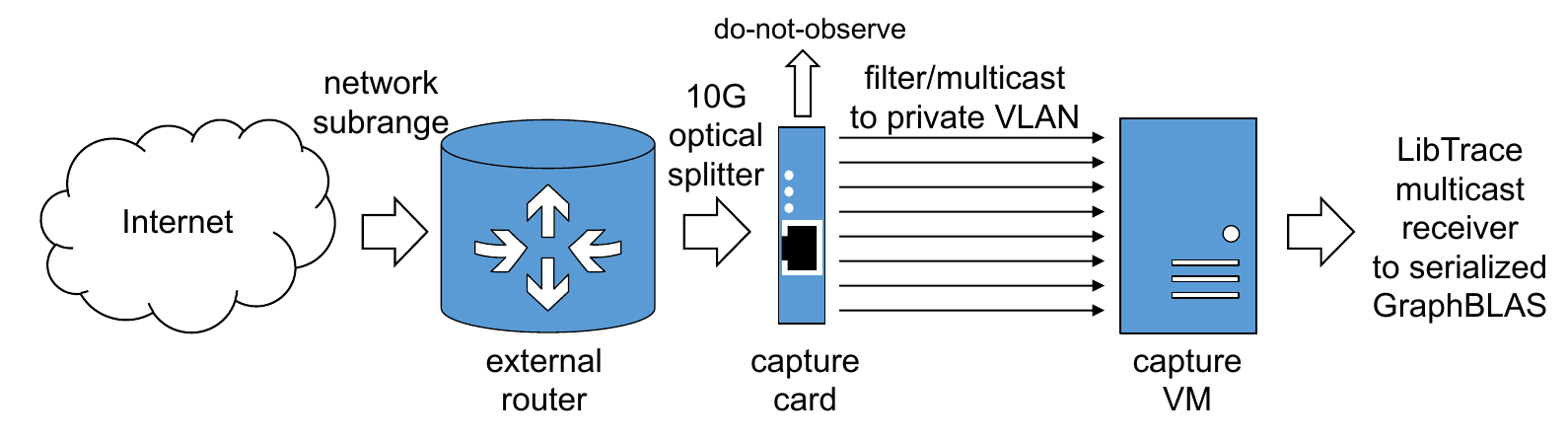}}
      	\caption{{\bf GraphBLAS Real-Time Deployment - Flow.} The Internet subrange is announced by the operator, and routed there, where an optical splitter routes traffic to a server containing an Endace DAG 10X2-P capture card which  filters out  do-not-observe  traffic before multicasting it onto a private VLAN.    LibTrace \cite{alcock2012libtrace} multicast clients on servers and VMs with interfaces on that private VLAN can subscribe to a multicast group and receive the raw feed  split up into 8 separate streams.}
      	\label{fig:GraphBLASflow}
\end{figure}

\begin{figure}
\center{\includegraphics[width=1.0\columnwidth]{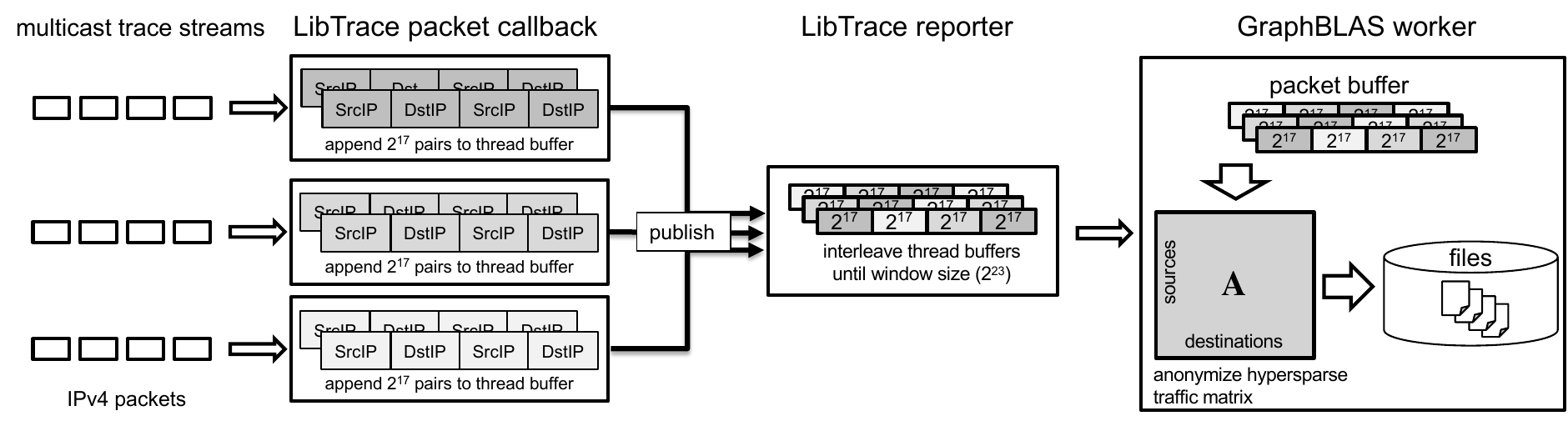}}
      	\caption{{\bf GraphBLAS Real-Time Deployment - LibTrace.} A program running in a VM on an analysis cloud  can use the LibTrace software to subscribe to a multicast group providing trace streams of the network traffic data.  The full  feed is broken up into 8 separate streams, which can be processed by separate threads in an application implementing the LibTrace API. The per-packet API callback is used to extract the source and destination IP addresses from each individually arriving packet and append them into a per-thread/per-stream packet buffer. When $2^{17}$ packets are received by one of these per-packet stream processing threads, it publishes this block to a single reporter thread, which appends newly-arriving blocks of $2^{17}$ packets together in a larger buffer.  After $2^{23}$ packets (64 published blocks) are received by the reporter thread, it loops through each element of this larger block, anonymizing each IP address before building (row, col, packet count) vectors out of them. The vectors for each set of $2^{17}$ packets are then used to build a $2^{32}{\times}2^{32}$ anonymized hypersparse GraphBLAS matrix.  The GraphBLAS matrices are serialized, compressed using Zstd level 1, and the result is saved to disk as 64 files inside a single UNIX TAR file representing $2^{23}$ packets.}
      	\label{fig:GraphBLASlibTrace}
\end{figure}

\section{GraphBLAS Real-Time Deployment}

     The data volumes, processing requirements, and privacy concerns of analyzing a significant fraction of the Internet have been prohibitive.  The North American Internet generates billions of non-video Internet packets each second \cite{Cisco2017, Cisco2018-2023}.   The GraphBLAS standard  provides significant performance and compression capabilities which improve the feasibility of analyzing these volumes of data.  Specifically, the GraphBLAS is ideally suited for both constructing and analyzing anonymized hypersparse traffic matrices.  Prior work with the GraphBLAS has demonstrated rates of 200 billion hypersparse matrix entries per second on a supercomputer \cite{kepner2021vertical}, while achieving compressions of 1 bit per packet \cite{trigg2022hypersparse}, and enabling the analysis of the largest publicly available historical archives with over 40 trillion packets \cite{kepner2021spatial}.

It should be noted, that GraphBLAS anonymized hypersparse traffic matrices represent only one set of design choices for analyzing network traffic.  Specifically, the use case requiring some data on all packets (no down-sampling), high performance, high compression,  matrix-based analysis, anonymization, and open standards.  There are a wide range of alternative graph/network analysis technologies and many good implementations  achieve performance close to the limits of the underlying computing hardware  \cite{tumeo2010efficient, kumar2018ibm, ezick2019combining, gera2020traversing, azad2020evaluation, du2021interactive, acer2021exagraph, blanco2021delayed, ahmed2021online, azad2021combinatorial, koutra2021power}.  Likewise, there are many network analysis tools that focus on providing a rich interface to the full diversity of data found in network traffic \cite{hofstede2014flow, sommer2003bro}.  Each of these technologies has appropriate use cases in the broad field of Internet traffic analysis.

The operational implementation described here builds on prior work \cite{jones2022graphblas} that demonstrated the ability of GraphBLAS to construct anonymized hypersparse traffic matrices at rates consistent with terabit networks using a few processing cores.  The mathematical functionality is briefly summarized as follows.

\subsection{Mathematics and Anonymization}

An essential aspect of this implementation is the use of constant packet, variable time sample windows, each with the same number of events/packets (denoted $N_V$).  Network traffic is dynamic and exhibits varying behavior on a wide range of time scales.  A given packet window size $N_V$ will be sensitive to phenomena on its corresponding timescale.  Determining how network quantities scale with $N_V$ provides insight into the temporal behavior of network traffic.  Constant packet, variable time samples  simplify the statistical analysis of the heavy-tail distributions commonly found in network traffic quantities \cite{nair2020fundamentals}. The contiguous nature of these data allows the exploration of a wide range of packet windows typically from $N_V = 2^{17}$ (sub-second) to $N_V = 2^{27}$ (minutes), providing a unique view into how network quantities depend upon time.  Efficient computation of network quantities on multiple time scales can be achieved by hierarchically aggregating data in different time windows \cite{trigg2022hypersparse}.

Internet data must be handled with care requiring trusted data sharing best practices that combine anonymizing source and destinations  with data sharing agreements. These data sharing best practices are the basis of the architecture presented here and include the following principles  \cite{kepner2021zero} 
\begin{itemize}
\item Data is made available in curated repositories
\item Using standard anonymization methods where needed: hashing, sampling, and/or simulation
\item Registration with a repository and demonstration of legitimate research need
\item Recipients legally agree to neither repost a corpus nor deanonymize data
\item Recipients can publish analysis and data examples necessary to review research
\item Recipients agree to cite the repository and provide publications back to the repository
\item Repositories can curate enriched products developed by researchers
\end{itemize}
Collection at the network source allows the data owner to construct and own the anonymization scheme and only share anonymized data under trusted data sharing agreements with the parties tasked with analyzing the data \cite{pisharody2021realizing}.

One of the important capabilities of the SuiteSparse GraphBLAS library is efficient support of hypersparse matrices where the number of nonzero entries is significantly less than either dimensions of the matrix.  If the packet source and destination identifiers are drawn from a large numeric range, such as those used in the Internet protocol, then a hypersparse representation of ${\bf A}_t$ eliminates the need to keep track of additional indices and can significantly accelerate the computations \cite{trigg2022hypersparse}.

Because matrix operations are generally invariant to permutation (reordering of the rows and columns), these quantities can readily be computed from anonymized data.  Furthermore, the anonymized data can be analyzed by subranges represented as subsets of IPs using simple matrix multiplication.  For a given subrange represented by an anonymized hypersparse diagonal matrix ${\bf A}_r$, where ${\bf A}_r(i,i) = 1$ implies  source/destination $i$ is in the range, the traffic within the subrange can be computed via: ${\bf A}_r {\bf A}_t  {\bf A}_r$. Likewise, for additional privacy guarantees that can be implemented at the  edge, the same method can be used to exclude a range of data from the traffic matrix
$$
     {\bf A}_t - {\bf A}_r {\bf A}_t  {\bf A}_r
$$

\begin{figure}
\center{\includegraphics[width=1.0\columnwidth]{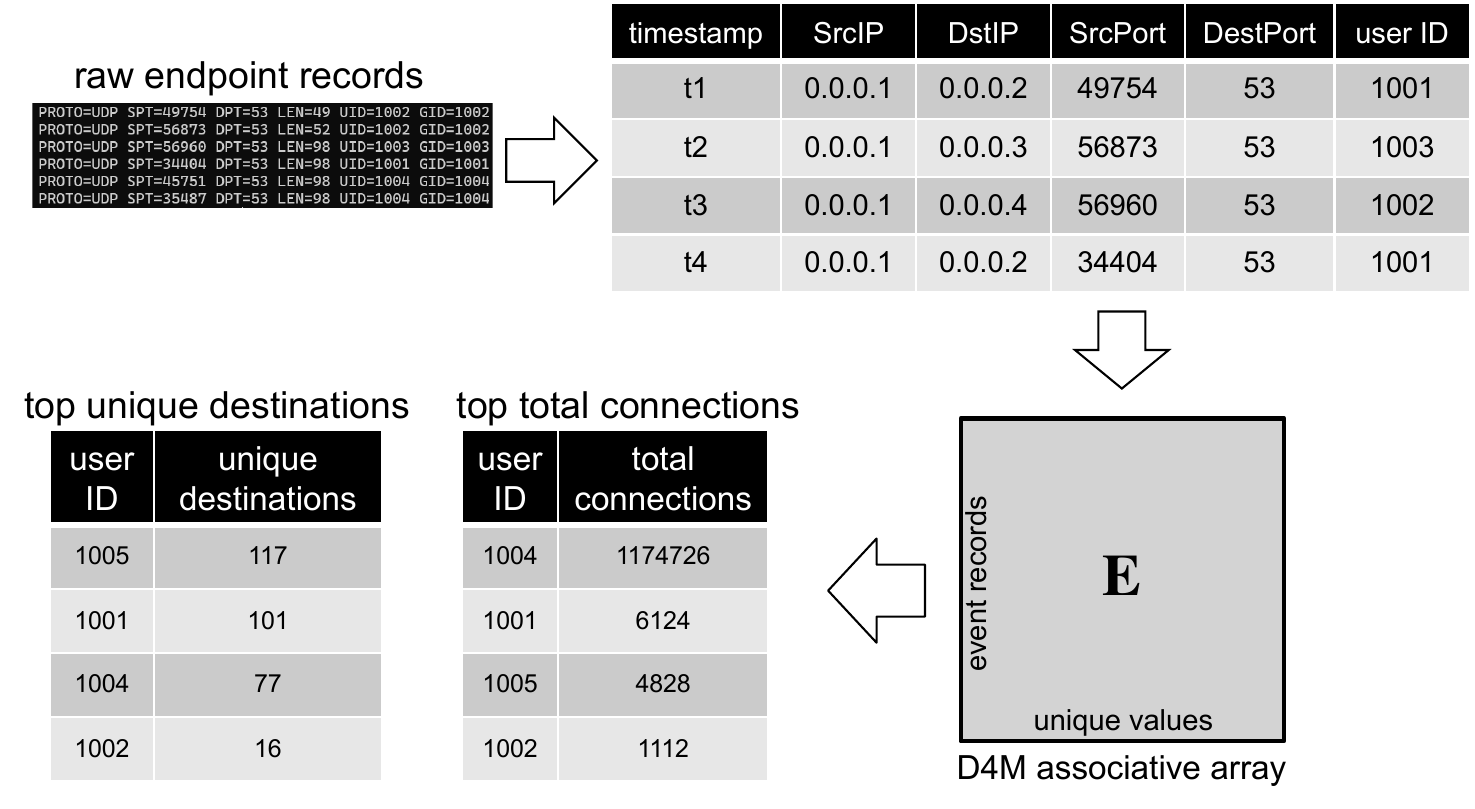}}
      	\caption{{\bf D4M Real-Time Deployment.} Internet-connected hosts on a compute cloud have netfilter rules which log the details of all newly originated IP sessions with external (Internet) destinations. The logs include the user ID of the connection’s owner.  These connection details are forwarded to a central log server in near real-time where they are logged to disk, and a nightly cron job parses them into tabular (TSV) format.   A secondary process reads this tabular data into a D4M  environment, converts them to associative arrays, and performs analytics to produce a daily report.}
      	\label{fig:D4Mflow}
\end{figure}

\subsection{Implementation}

The operational implementation of the above functionality is depicted in Figure~\ref{fig:GraphBLASflow} and Figure~\ref{fig:GraphBLASlibTrace}.  In this implementation an Internet subrange is announced by the operator, and routed to the operator, where an optical splitter routes traffic to a server containing an Endace DAG 10X2-P capture card which  filters out do-not-observe traffic before multicasting it onto a private VLAN (virtual local area network).    The LibTrace library \cite{alcock2012libtrace} multicasts the stream to clients running on servers in VMs (virtual machines) that have interfaces on the private VLAN.  These clients can subscribe to a multicast group and receive the raw feed  round-robin split  into 8 separate streams to maintain performance.

Inside the clients, the per-packet LibTraceAPI callback is used to extract the source and destination of 32-bit IPv4 addresses from each individually arriving packet and append them into a per-thread/per-stream packet buffer. When $2^{17}$ packets have been received by one of these per-packet stream processing threads, it publishes this block to a single reporter thread, which appends newly-arriving blocks of $2^{17}$ packets together in a larger buffer.  After $2^{23}$ packets (64 published blocks) have been received by the reporter thread, it loops through each element of this larger block, anonymizing each IP address before building (row, col, packet count) vectors out of them.  The anonymization of the source and destinations is performed using CryptoPAN \cite{fan2004prefix} either directly or via a pre-populated $2^{32}$ element lookup table.  This is a simple size-speed trade-off.  If speed is important, and a lookup table can fit in memory, the lookup table is faster.  If memory is important,  CryptoPAN  can be run directly on the data, resulting in an  $\sim$100x less memory usage at the cost of a  $\sim$10x slowdown.  Other anonymization schemes can be chosen that offer different trade-offs \cite{dandyan2022feistel}.  The anonymized vectors for each set of $2^{17}$ packets is then used to build a $2^{32}{\times}2^{32}$ anonymized hypersparse GraphBLAS matrix.  The GraphBLAS matrices are serialized, compressed using Zstd level 1, and the result is saved to disk as 64 files inside a single UNIX TAR file representing $2^{23}$ packets.  The TAR files are then transmitted to another system for archiving and more in-depth off-line analysis.  The time to save compress and save to disk is negligible and not a factor in overall performance.

\section{D4M Real-Time Deployment}

The D4M library (d4m.mit.edu) implements associative array mathematics in multiple languages (Python, Julia, Matlab, and Octave) \cite{chen2016julia, milechin2018d4m, gadepally2018hyperscaling, jananthan2022pyd4m}.  As described earlier, associative arrays are a generalization of matrices that allow unstructured data to be used as rows, columns, or values.  D4M is particularly useful in analyzing log data or interfacing with databases.  Many firewalls and cloud systems maintain raw connection logs to support operational capability and provide visibility into potentially unwanted behavior on or by the system.  These connection logs often contain diverse unstructured data.  Figure~\ref{fig:D4Mflow} shows an operational example of how D4M is used to analyze the daily connections logs of a cloud system.  Internet-connected hosts on a compute cloud have netfilter rules that log the details of all newly originated sessions with external (Internet) destinations. The logs include the user ID of the connection’s owner.  These connection details are forwarded to a central log server in near real-time where they are logged to disk, and a nightly cron job parses them into to tabular (TSV) format.   A secondary process reads this tabular data into a D4M  analysis environment, converts them to associative arrays, and performs analytics to produce a daily report

Mathematically, the tabular log data is transformed to a associative array ${\bf E}$ where the rows are the event records and the columns are unique values of the record.  Such an event array tends to be very sparse since each unique value has is a column.  Tallying records can then be done easily with array multiplication.  For example,  creating a user ID by destination array is done be multiplying sub-arrays holding the appropriate fields of data
$$
    {\bf A}_{{\rm userID}{\times}{\rm DstIP}} = {\bf E}_{\rm userID}^{\sf T} ~  {\bf E}_{\rm DstIP}
$$
Top connection counts and destination counts by userID can be calculated by applying the appropriate formulas from Table~\ref{tab:Aggregates}
$$
      {\bf A}_{{\rm userID}{\times}{\rm DstIP}} {\bf 1} ~~~~~  {\rm and} ~~~~~  |{\bf A}_{{\rm userID}{\times}{\rm DstIP}}|_0  {\bf 1}
$$ 
Likewise, top connections and userID  by  destination can be calculated by multiplying by  ${\bf 1}^\top$ on the left instead of the right
$$
      {\bf 1}^{\sf T} {\bf A}_{{\rm userID}{\times}{\rm DstIP}}  ~~~~~  {\rm and} ~~~~~ {\bf 1}^{\sf T} |{\bf A}_{{\rm userID}{\times}{\rm DstIP}}|_0 
$$

\section{Results}

\begin{figure}
\center{\includegraphics[width=1.0\columnwidth]{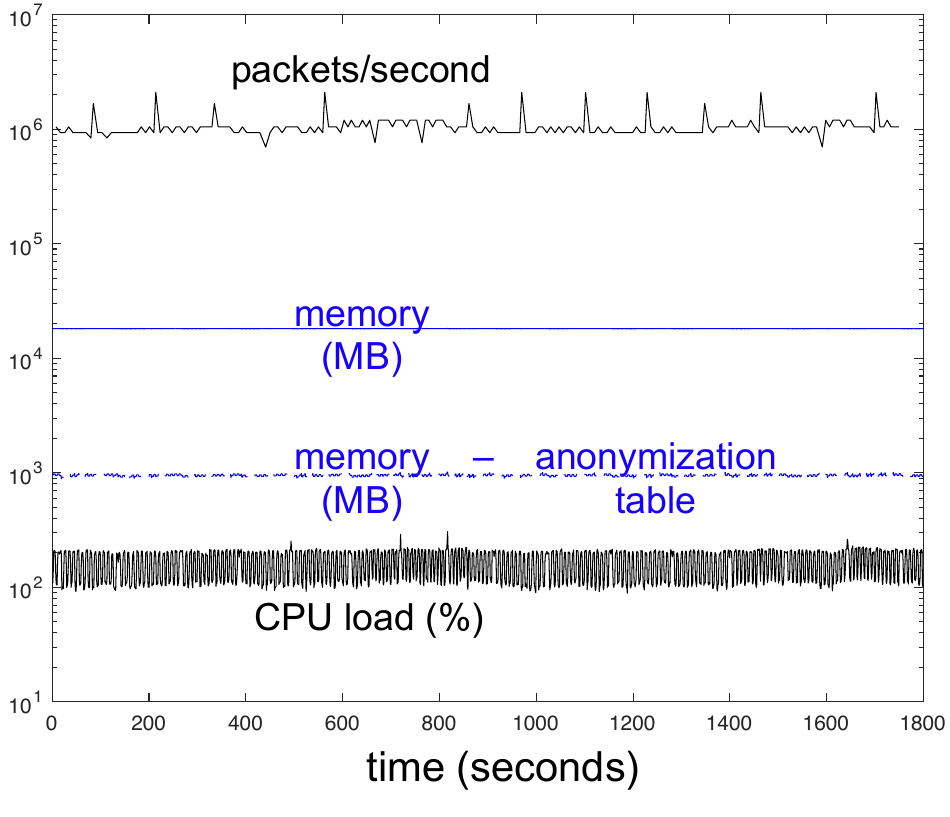}}
      	\caption{{\bf GraphBLAS Results - Overview.}  Packets/second processed, total memory, active memory, and processing load of over a 30 minute monitoring period (in the midst of a several month run) for the GraphBLAS real-time deployment showing the stability of the processing pipeline during this period.}
      	\label{fig:GraphBLAScapturePlot}
\end{figure}

\begin{figure*}
\center{
\includegraphics[width=0.97\columnwidth]{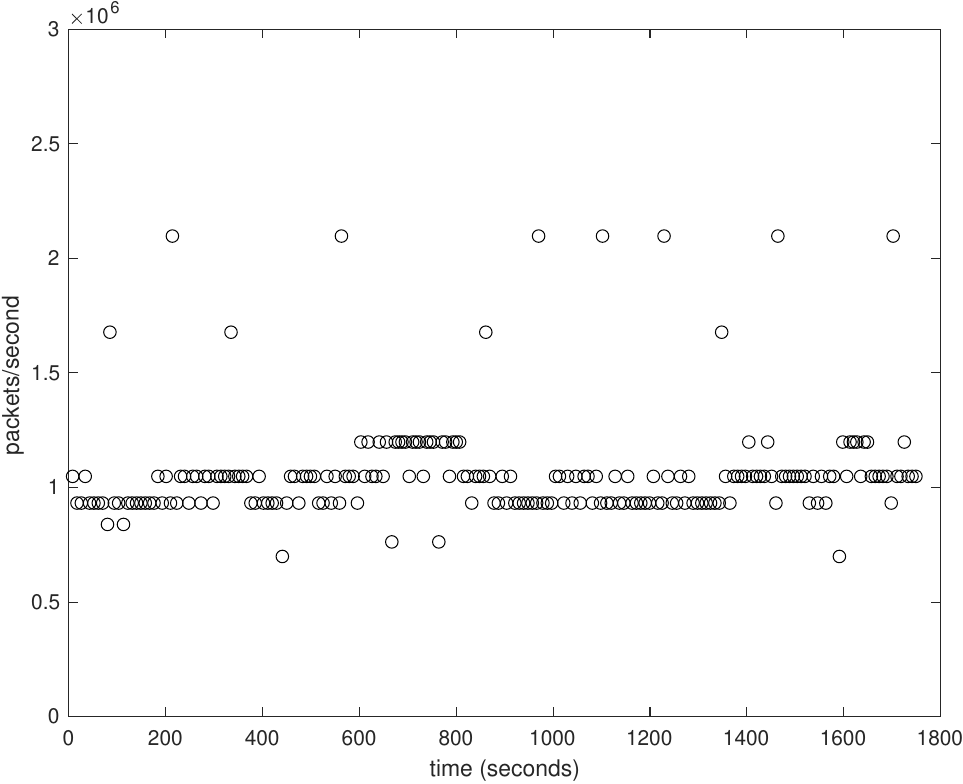}
\includegraphics[width=1.0\columnwidth]{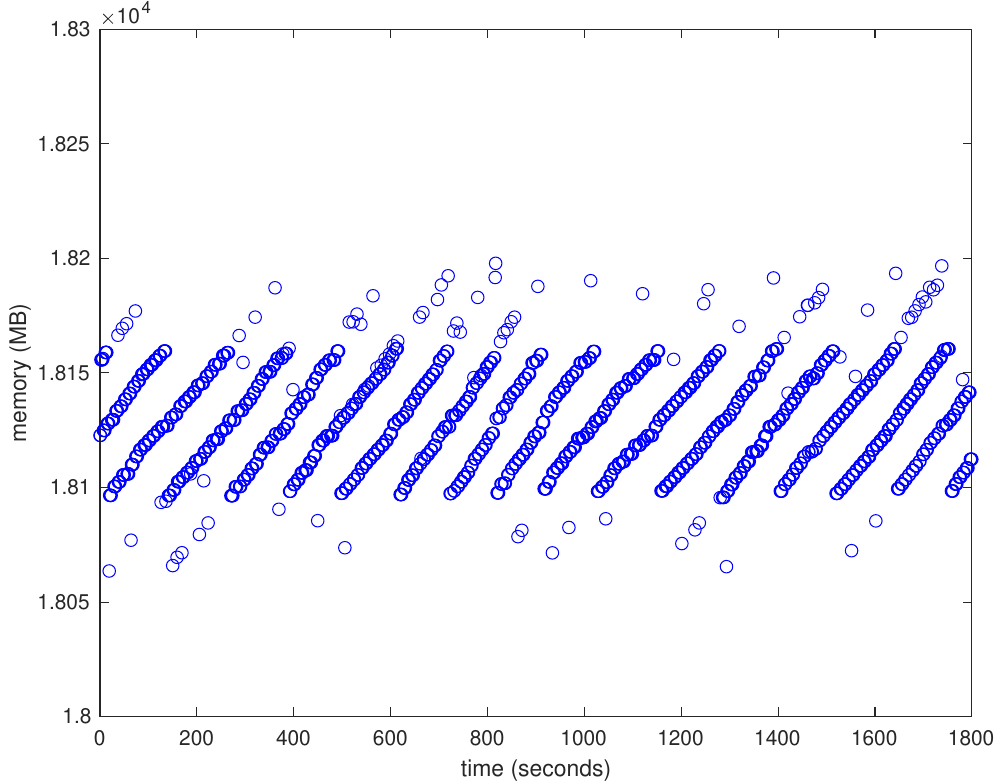}
\includegraphics[width=0.97\columnwidth]{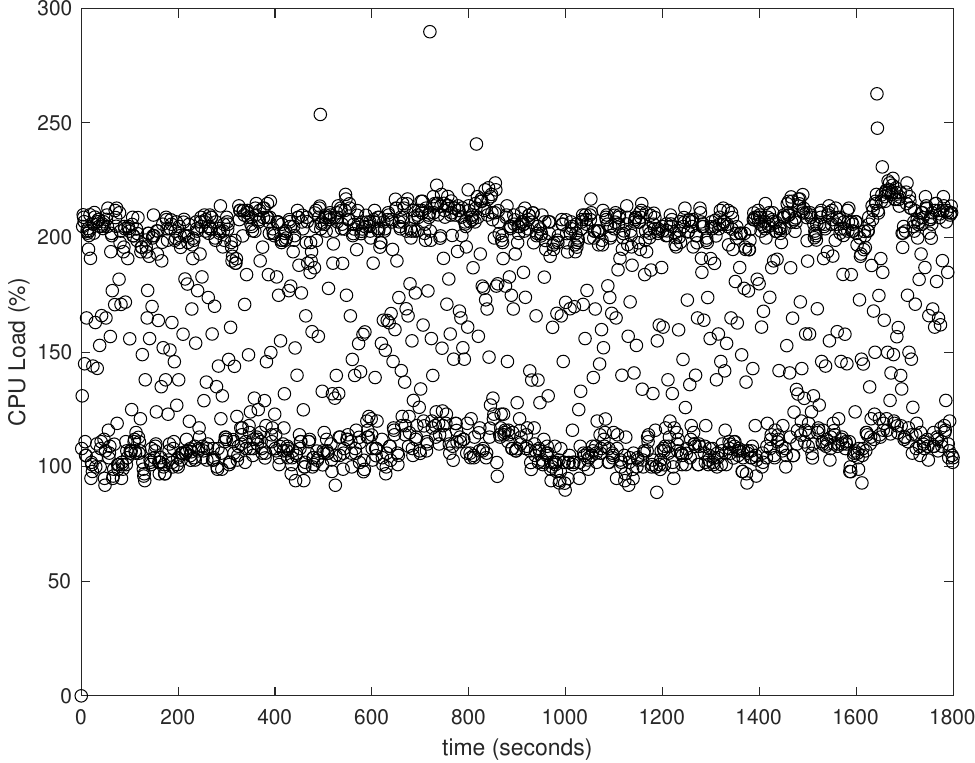}
\includegraphics[width=0.98\columnwidth]{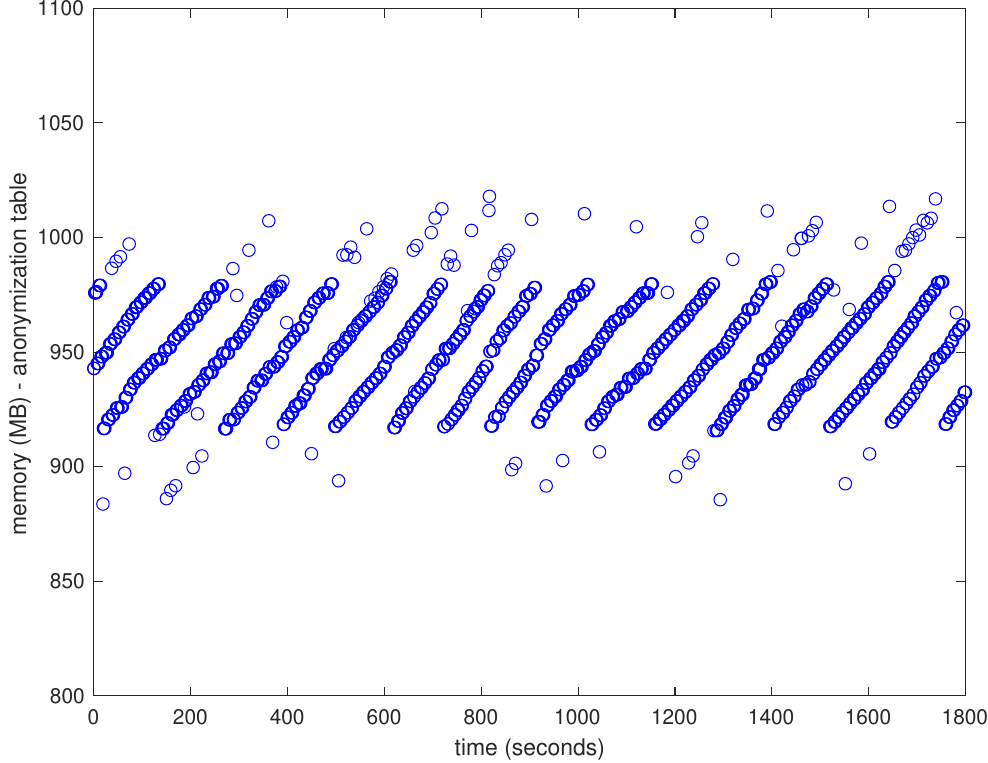}
}
      	\caption{{\bf GraphBLAS Results - Detail.}  (top left) Packets/second varies from $7{\times}10^5$ to $2.1{\times}10^6$.  The discrete levels are a result of the 1 second granularity of the timer. (bottom left) CPU load as a percentage of core utilization. The load varies distinctly from fully using one (100\%) or two (200\%) cores. (top right) Total memory used.  (bottom right)  Total memory used minus the 16 GB anonymization table. The memory fluctuations are consistent with memory allocation requirements of the packet buffers.}
      	\label{fig:GraphBLASplot4}
\end{figure*}

The goal of this work is to illustrate some effective architectures for deploying matrix/array-based analysis on streaming networks and to set the expectations on the potential rates and resources required.  For both of the implementations measurements were taken of their rates of execution and resource consumption during operation.

\subsection{GraphBLAS Real-Time Deployment}

Performance measurements for the GraphBLAS real-time deployment were collected on  a 8-core VM with 32 GB of RAM hosted on a research compute cloud.  Measurements were collected at 1-second intervals for 30 minutes with {\sf psrecord}, a small Python utility that uses the {\sf psutil} library to record the CPU and memory activity of a process by polling statistics from the Linux {\sf /proc} filesystem, using the command
\begin{verbatim}

psrecord --interval 1 --duration 1800
         --log psrecord-real.txt <PID>
\end{verbatim}
In addition, the timestamps of each TAR file created were recorded.  These data were collected on an instance of the program that had been running for many months without interruption and represent a good example of the sustained resource requirements for the capability.

Figure~\ref{fig:GraphBLAScapturePlot} presents and overview of rates, total memory, active memory, and processing load over the time of the collection.  The rates are computed by counting the number events/packets in a file and dividing that by the differences in the time stamps between successive files.   The active memory is the total memory minus the 17 GB CrypoPAN lookup table.  Each of the quantities is relatively stable over the collection period. Figure~\ref{fig:GraphBLASplot4} provides more detailed looks at each of the measured quantities.  Packets/second varies from $7{\times}10^5$ to $2.1{\times}10^6$.  The discrete levels are a result of the 1 second granularity of the timer. CPU load is shown as a percentage of core utilization. The CPU load varies distinctly from fully using one (100\%) or two (200\%) cores.  Profiling indicates that the dominant operation is the GraphBLAS hypersparse matrix constructor.  The observed load  is consistent with  8 threads running on the 8 available cores with a maximum effective GraphBLAS hypersparse matrix constructor duty cycle of 25\%.  The memory fluctuations are consistent with memory allocation requirements of the packet buffers and the natural fluctuations of the underlying LibTrace buffers as it reacts to spikes in packet volume over time.  These memory requests are $<$0.1\% of the peak memory bandwidth and have a negligible impact on performance.

\subsection{D4M Real-Time Deployment}

\begin{figure}
\center{\includegraphics[width=1.0\columnwidth]{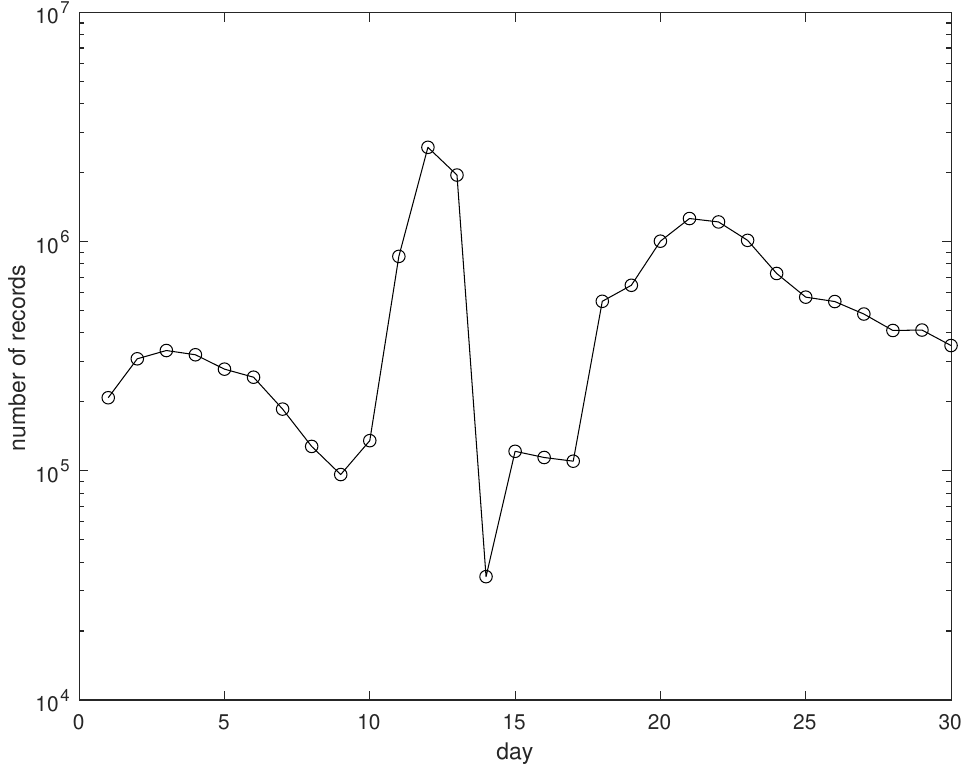}}
      	\caption{{\bf D4M Records Collected per Day.}  Number of records (distinct IP sessions) recorded per day, which varies significantly between $3{\times}10^4$ and $3{\times}10^6$ records per day.}
      	\label{fig:D4MnRecord-crop}
\end{figure}

\begin{figure}
\center{\includegraphics[width=1.0\columnwidth]{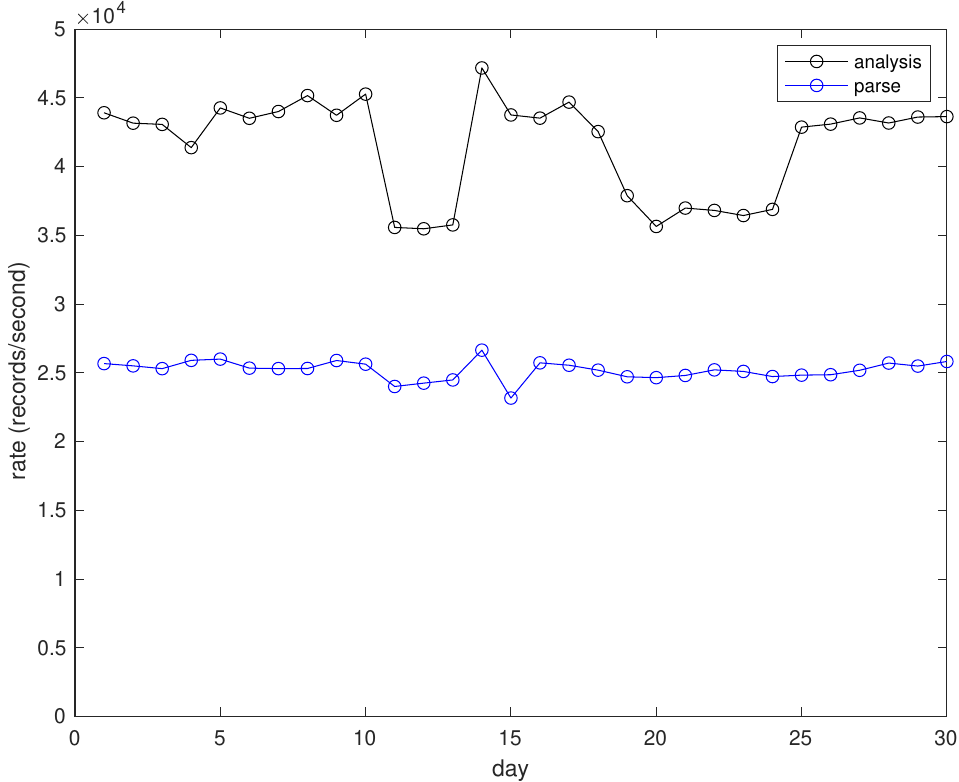}}
      	\caption{{\bf D4M Parse and Analysis Rates.}  Parse and analysis rates in terms of records per second.  Parse rates are typically $2.5{\times}10^4$  per second and analysis rates are in the $4{\times}10^4$ per second range.}
      	\label{fig:D4Mrate-crop}
\end{figure}

Performance measurements for the D4M analysis were taken using a dual Intel Xeon Platinum 8260 server with 192GB of RAM.  A single-process single-threaded instance of D4M running GNU Octave version 6.3.0 was used.  The execution of the parse and analysis steps were timed separately over 30 days of daily log analysis.  Figure~\ref{fig:D4MnRecord-crop} shows the number of connection records that were logged each day, which varies significantly between $3{\times}10^4$ and $3{\times}10^6$ records per day. Figure~\ref{fig:D4Mrate-crop} shows the parse and analysis rates in terms of records per second.  These are consistent with benchmarked single-process single-thread performance rates for the D4M analysis framework \cite{chen2016julia, gadepally2018hyperscaling, jananthan2022pyd4m} which are set by underlying performance of their language provided variable length string sorters.

\section{Conclusions and Future Work}

Significant insight into the behavior, operation, and protection  of networks can be gained via matrix/array-based analyses techniques.  This paper builds on prior work on GraphBLAS (graphblas.org) hypersparse matrices and D4M (d4m.mit.edu) associative arrays (a mathematical superset of matrices) by integrating them into two operational systems.  First, an operational GraphBLAS implementation that constructs anonymized hypersparse matrices on a high-bandwidth network tap.  Second, an operational D4M implementation that analyzes daily cloud gateway logs. The architectures of these implementations can be viewed as starting points for others trying to deploy matrix/array-based network analyses techniques.  Detailed measures of the resources and performance of these implementations indicate they are capable of keeping up with their operational requirements using modest computational resources (a couple of processing cores).   GraphBLAS is well-suited for low-level analysis of high-bandwidth connections with relatively structured network data.  D4M is well-suited for higher-level analysis of more unstructured data.

Future work in this area could expand the ecosystem for matrix/array-based network analyses by incorporating these methods into next generation cloud operating systems, programmable network technologies \cite{burstein2021nvidia,burres2022intel}, the hierarchical integration of low-level network with high-level network data, and privacy-preserving network protection strategies.

\section*{Acknowledgments}

The authors wish to acknowledge the following individuals for their contributions and support: Daniel Andersen, Sean Atkins, Chris Birardi, Bob Bond, Andy Bowne, Stephen Buckley, K Claffy, Cary Conrad, Chris Demchak, Alan Edelman, Garry Floyd, Jeff Gottschalk, Dhruv Gupta, Chris Hill, Kurt Keville, Charles Leiserson, Chase Milner, Sanjeev Mohindra,  Dave Martinez, Joseph McDonald, Sandy Pentland, Heidi Perry, Christian Prothmann, John Radovan, Steve Rejto, Josh Rountree, Daniela Rus, Scott Weed, Marc Zissman.

\bibliographystyle{ieeetr}
\bibliography{GraphBLAS-D4M-Deploy}

\end{document}